\documentclass[prepreint,aip, ajp, nofootinbib]{revtex4-2}
\usepackage{graphicx}
\usepackage{amsmath}

\usepackage{siunitx}
\usepackage{booktabs}
\usepackage{esvect}

\usepackage{color, soul}

\begin{document}

\title{Computational Physics in the Advanced Lab: Experiment and Simulation of Thermal Diffusion in Metal Rods}

\author{Yash Mohod}

\author{M.\ C.\ Sullivan}
\email{mcsullivan@ithaca.edu}

\affiliation{Department of Physics \& Astronomy, Ithaca College, Ithaca, New York 14850}

\date{\today}

\begin{abstract}
We present a simple experiment on thermal diffusion in metal rods that can be used to integrate computation into the advanced lab.  An analytical solution exists for the transient heat conduction in an infinite rod with a delta function heat input, but no analytical solution exists for short rods or for long duration heat inputs. Our apparatus is a copper rod with a heater and thermometers attached to the rod.  The temperature of the metal rod due to transient heat conduction can be modeled using a simple numerical simulation and the finite centered difference method.  Using a 22 cm long copper rod with the ends thermally sunk in aluminum blocks, we show poor agreement between the experimental results and the infinite-rod analytical model, but excellent agreement between the experimental results and our numerical simulation.  This experiment shows the power of a numerical simulation but also the limitations of the chosen model, which can be used as motivation for further exploration.

\end{abstract}

\maketitle

\section{Introduction}
Since their invention, computers have been integral to scientific progress, from the first computers used during the Manhattan Project\cite{manhattan} to the most recent 1000+ qubit quantum computers.\cite{wikipedia_QuantumComp}  In the mid-1990s, many physics departments added computational physics to their undergraduate curricula, often as a standalone course.  Reflecting this change, the American Journal of Physics produced the first computational physics Resource Letter in 1996,\cite{CP1} updated it in 2008\cite{CP2} and again in 2023.\cite{CP3}  In many programs, computational physics is now integrated into multiple courses throughout the curriculum, as evidenced by the work of organizations like PICUP: The Partnership for Integration of Computation into Undergraduate Physics.\cite{picup}

Surprisingly, despite obvious connections, computational physics has been slow to integrate into advanced laboratory courses.  There are no advanced laboratory resources mentioned in any computational physics resource letter.\cite{CP1, CP2, CP3} In the first (and only) advanced laboratory Resource Letter,\cite{ALC1} only seven (out of 91) resources involved computational physics.  This article expands the available resources for instructors who wish to integrate computational physics into the advanced laboratory by providing a simple experiment in thermal diffusion whose results cannot be predicted analytically, only numerically.

Thermal diffusion is fundamental to the study of thermodynamics, and there are multiple textbooks to describe the subject for physicists and engineers.\cite{schroeder, textbook}  In addition, the thermal diffusion equation is a gateway into the study of partial differential equations, representing the next step beyond the harmonic oscillator and exponential growth/decay.  Previous experiments in thermal diffusion compared experimental results to analytical models with heat flow in one or two dimensions.\cite{Sullivan_AJP_2008, Brody_AJP_2017, Gfroerer_AJP_2015} Only one recent experiment compared experimental results to a numerical simulation,\cite{McDougall_AJP_2014} achieving a roughly 1\% deviation between the experiment and simulation.

In this paper we explore thermal diffusion in copper rods using a geometry that is experimentally accessible and that showcases the power of computational physics.  We relax the experimental constraints that lead to an analytical solution of the differential equation and explore the effects of the rod length, heat pulse duration, and different boundary conditions, and we introduce a flexible numerical simulation that gives us the ability to explore these more complex experimental situations. These situations cannot be solved analytically, but the numerical model can be readily verified via experiment.

\section{Apparatus and Procedure}
\begin{figure}
    \fbox{\includegraphics[width=0.8\linewidth]{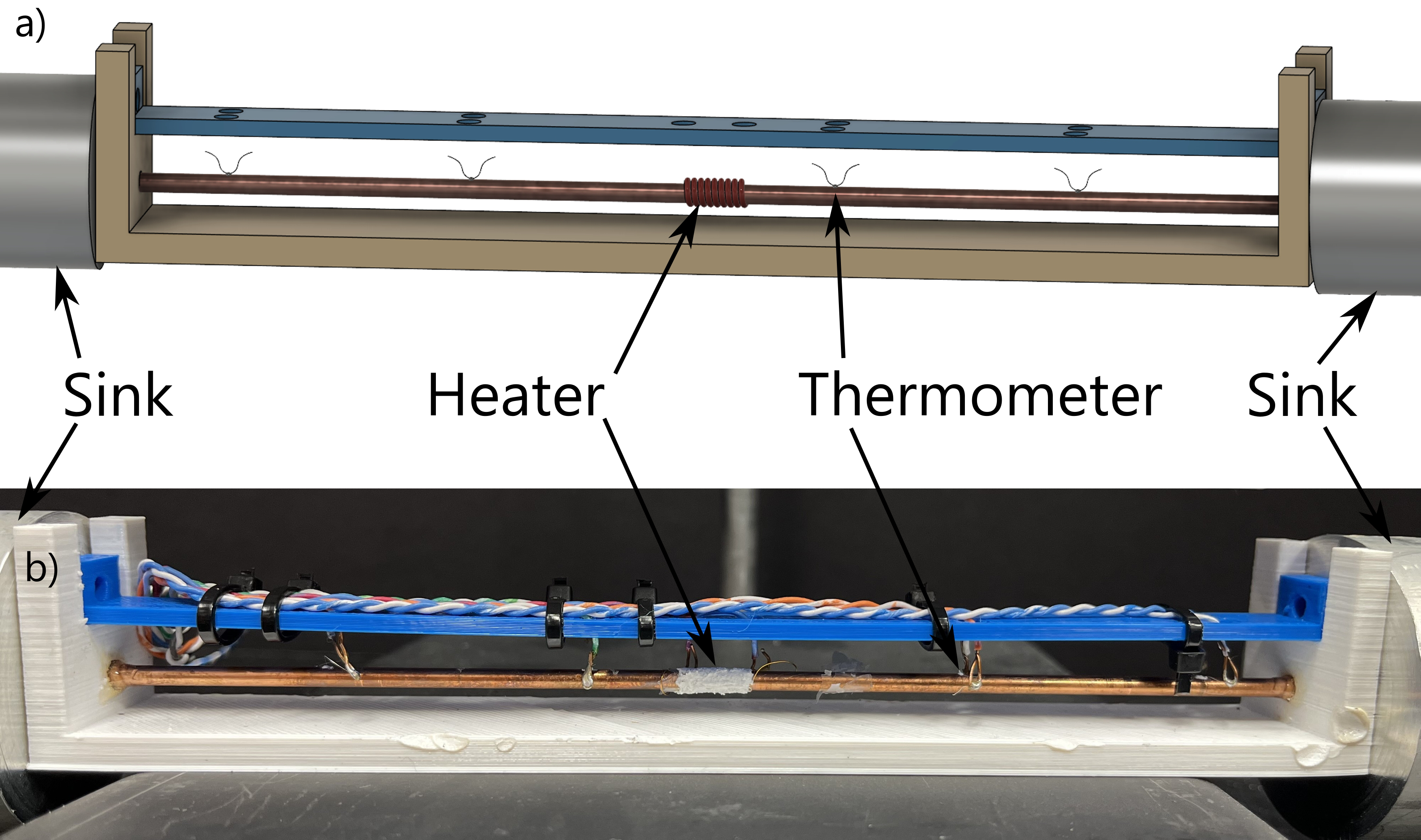}}
    \caption{(Color online) The experimental apparatus, consisting of a copper rod with aluminum sinks on either end, heater in the center, and four thermistors to measure temperature changes along the rod.  The rod is supported by a 3D printed white plastic frame, with a blue bridge above the rod to hold the connection wires.  Not pictured is a cover to fully enclose the rod and wire connections.  Panel (a) shows a 3D render of the apparatus, and (b) is a photograph of the apparatus.}
    \label{fig:apparatus}
\end{figure}

The apparatus and its frame are shown in Fig.\ \ref{fig:apparatus}.  Our apparatus, based on the one discussed in Ref.\ \citenum{Sullivan_AJP_2008}, consists of a copper rod 22 cm long and 3.2 mm in diameter, threaded 1 cm on each end to accept aluminum thermal sinks (discussed in detail in Sec.\ \ref{Al_sinks}).  In the center of the rod we wrapped about 50 turns of 0.16 mm diameter (34 gauge) coated phosphor bronze wire directly around the rod to act as a resistive heater (R = 3.2 $\Omega$).  These turns of wire covered a length of 1 cm of the rod and were held in place using thermally conductive epoxy. Fast-response thermistors (Honeywell 111-202CAK-H01) were placed at distances of 2 cm and 6 cm from the center of the rod on one side of the heater and at distances of 4 and 8 cm from the center on the other side of the heater.  This placement allows for multiple distances to be be read simultaneously while still leaving space for wiring.  

We designed and 3D printed a frame for the experimental apparatus, shown in white in Fig.\ \ref{fig:apparatus}.  The frame supports the rod and keeps the rod from twisting.  The twisting rotates the thermistors away from the electrical contacts and can break the thermistor wires (which are 0.03 mm in diameter and quite fragile).  The blue 3D printed ``bridge" above the rod helps to keep the electrical wiring from moving and applying stresses to the thermistor wires.  The open box shape was designed to allow access to the experimental apparatus, but we also designed a cover to fully enclose the experiment (not shown in Fig.\ \ref{fig:apparatus}).  Enclosing the apparatus allows us to pick up and move the experiment without fear of damaging the heater or thermistors.  The cover was in place for all the experiments. The 3D printed frame is 2 cm shorter than the rod to allow the attachment of aluminum sinks on both ends of the copper rod.\cite{EPAPS}

To create temperature changes in the rod, we applied a current pulse to the heater coil that generated a heat pulse via Joule heating.  The subsequent temperature change at different locations on the rod was measured at each thermistor by using a voltage divider and an op-amp amplifier.  The details of the data collection system are described in Ref.\ \citenum{Sullivan_AJP_2008}.\footnote{The only difference between our data collection system and the one described in Ref.\ \citenum{Sullivan_AJP_2008} is that we measured the gain of each amplifier circuit to find $G = 20$ (predicted gain $G=21$), and we use the exact formula to convert measured voltage difference ($dV_A$) to temperature difference ($dT_{th}$): $dT_{th} = A\, T_a^2\, dV_A/(1-A\, T_a \,dV_A)$, where $A = 4/(G\,T_0\,V_0)$, and $T_a$ is the ambient temperature before and after measurement, $T_0$ is the temperature of the semiconductor gap ($T_0 =3068$ K for this thermistor), and $V_0$ is the reference voltage for the amplifier circuit ($V_0 = 5$ V).}  To reduce heat loss via convection, the entire apparatus was placed in a bell jar and evacuated to about $P \approx 1\times10^{-5}$ mbar (0.001 Pa).  We take 100 s of data before turning on the heater to measure the thermal drift in the system.  We fit a line to the thermal drift and assumed the same linear behavior for the whole data run, and then subtracted this thermal drift from our measurements.  Each measurement took 300-600 s, and we waited 300 - 600 s between each measurement.  This yielded 700 - 1300 s between each heat pulse to allow the system to come to a new thermal equilibrium.

\section{Thermal Diffusion: Model and Solutions} \label{sec:DiffusionModel}

Along a metal rod with heat flow in one dimension, the thermal diffusion equation takes the form:

\begin{equation}
s \frac{\partial \Theta}{\partial t} = k \frac{\partial
^2\Theta}{\partial z^2} + v - w\Theta.\label{eq:heat-flow-pde-2}
\end{equation}
Here, $z$ is the location along the rod, $\Theta$ is the temperature difference between temperature of the surroundings (``ambient" temperature) and the temperature of the rod ($\Theta(z,t) = T(z,t)-T_a$), $k$ is the thermal conductivity of the rod, and  $s$ is its volumetric heat capacity $s = c_P\, \rho$, where $c_P$ is the specific heat at constant pressure and $\rho$ is the material density.  The first two terms -- the partial derivatives -- are the diffusion equation, and explain how the temperature difference diffuses in position $z$ as a function of time $t$.  The next term $v$, where $v=v(z,t)$ is the power density input for the rod. This allows us to add energy to the rod and increase its temperature.   


The last term represents heat loss on the surface of the rod, which can be due to both convection and radiation.  Convective heat loss, commonly called Newton's law of cooling, is proportional to the temperature difference $\Theta$ between the rod and the gas surrounding the rod.  Radiative heat loss depends on the temperature of surroundings, in our case, the frame and cover of the bell jar.  If we assume the surroundings are the same temperature as the gas then the radiative heat loss is proportional to ($T^4 - T_a^4$).  This difference can factored as a quartic polynomial in $\Theta$, where for small temperature differences only the linear term is significant.\cite{OSullivan_AJP_1990}  Researchers have shown good agreement with a linear radiative heat loss for temperature differences as large as 40 K,\cite{Vollmer_EJP_2009} much larger than the temperature differences ($\Theta_{max}\approx 1-5$ K) in this experiment.
 
For our small temperature changes, $w = 2 h/a$, where $a$ is the radius of the rod and $h$ is the heat transfer coefficient.  We know $h = h_c + 4 \epsilon \sigma T_a^3$, where $\sigma$ is the Stefan–Boltzmann constant ($5.67\times10^{-8}$ Wm$^{-2}$ K$^{-4}$), $\epsilon$ is the surface emissivity, and $h_c$ is the convective heat transfer coefficient.  The convective heat transfer coefficient typically varies between 2 and 25 Wm$^{-2}$ K$^{-1}$ at atmospheric pressure.\cite{Vollmer_EJP_2009, Sullivan_AJP_2008} In our low pressure setup, we expect $h_c \approx 0$.\cite{lowpressure}  If the ambient temperature is at room temperature, then the radiative factor $4 \sigma T_a^3 \approx 6$ Wm$^{-2}$ K$^{-1}$, although the emissivity $\epsilon$ for semi-polished metals is typically $\epsilon=0.2$ or lower,\cite{Vollmer_EJP_2009, Sullivan_AJP_2008} meaning the heat loss from radiation is expected to be of the order of 1 Wm$^{-2}$ K$^{-1}$.  We will fit our data assuming a linear heat loss and compare our fit with this rough prediction.

From the diffusion equation we can find a characteristic time as a function of the diffusion length, $\tau = z^2/D$, where the diffusion coefficient $D$ in a thermal system is $D = k/s$.  For a copper rod at $z=2$ cm, $\tau \approx 3$ s (for $z=8$ cm, $\tau \approx 55$ s).  This characteristic time sets the timescale for diffusion in our system.  For example, we expect to reach a steady state in our system at times that are tens to hundreds of times larger than the characteristic time.

\subsection{Analytical Solution}(()Eq.\ \ref{eq:heat-flow-pde-2}) if we make assumptions about the heater, the heat input $v(z,t)$ and the boundary conditions.  First, we assume that a total amount of energy $Q$ is delivered to the rod instantaneously.  Second, we assume the heater is infinitesimally small.  Finally, we assume the rod is infinitely long, which resolves the boundary conditions.  Under these assumptions, we can solve Eq.\ \ref{eq:heat-flow-pde-2} to find the temperature difference at all times and at all locations along the rod:\cite{Sullivan_AJP_2008}

\begin{equation}
\Theta(z,t) = \frac{Q}{2\pi a^2 \sqrt{\pi k s t}} \cdot e^{-z^2 s/4kt} \cdot
e^{-wt/s}. \label{eq:T-soln2}
\end{equation}

Previous work has shown good agreement between this analytical solution and experimental results using an 80 cm copper rod.\cite{Sullivan_AJP_2008}

\subsection{Numerical Model}
The heat flow equation (Eq.\ \ref{eq:heat-flow-pde-2}) can be solved for other assumptions and boundary conditions,\cite{Brody_AJP_2017} but there are many situations where the system is too complex to derive an analytical solution, and even simple systems can defy analytical solutions for certain boundary conditions.  For systems where the analytical solution is impossible, we can turn to the finite difference method, which approximates the rod by dividing it into small pieces.  In this paper, we will use the forward Euler method to approximate the time derivative in Eq.\ \ref{eq:heat-flow-pde-2} and the finite centered difference method to approximate the spatial derivative.\cite{Mazumder_2016}  Using this numerical model, we can now change the length of the rod, the size of the heater, the duration of the heat pulse, and the boundary conditions on the ends of the rod.

\subsubsection{Numerical Solutions}
The numerical model treats the rod as $N$ small sections of length $\Delta z$; the smaller the segment size, the more accurate the simulations are, but at the cost of higher computation time.  We can then solve for the temperature difference $\Theta_i$ of the $i$th segment using the finite centered difference method for a time step $\Delta t$:\cite{Mazumder_2016}
\begin{equation}
   \Theta_i(t+\Delta t) = \Theta_i(t) + \left(\frac{k(\Theta _{i+1}(t) - 2 \Theta _i(t) +\Theta_{i-1}(t))}{(\Delta z)^2} + v_i - w\Theta _i(t)\right)\cdot\frac{\Delta t}{s}. \label{eq:FCD}
\end{equation}
The finite centered difference transforms the thermal diffusion equation (Eq. \ref{eq:heat-flow-pde-2}) into the sum of the differences of the temperatures of the segment and its nearest neighbors.  Using this, we can model all $N$ segments of the rod from $t=0$ until $t=T$ , where $T$ is the total duration of the model. 

In Eq.\ \ref{eq:FCD}, the heat loss is represented as an array of values, $w \Theta _i$, associated with each segment $i$.  The value of heat loss parameter $w$ is unchanged.  

The heat input in Eqs.\ \ref{eq:heat-flow-pde-2} and \ref{eq:FCD} is given by the power density $v$. The power $P$ is the total power input to the heater; experimentally, we determine the power generated in the heater, $P = V^2/R$, by measuring the voltage $V$ across the heater and using the known resistance $R$ of the heater.  The length of the heater $L_{\mathrm{heat}}$ must be an integer number of rod segments.  If the radius of the rod is $a$, then the volume surrounded by the heater is $(\pi a^2 )L_{\mathrm{heat}}$.  If we assume the energy due to Joule heating is transferred instantaneously into the volume surrounded by the heater, then the power density $v_i$ per segment is given by:
\begin{equation}
   v_i = \frac{P}{(\pi a^2 )L_{\mathrm{heat}}}. \label{eq:v_i}
\end{equation}
We set $v_i = 0$ for all segments not under the heater, so no heat is added to those segments. For times after the heat pulse we set all values $v_i=0$.

Eq.\ \ref{eq:FCD} governs all rod segments, excluding the first ($i=0$) and last ($i=N$) segments of the rod.  The time evolution of these segments is determined by the boundary conditions. There are three commonly used boundary conditions when solving differential equations: Dirichlet (constant value), Neumann (constant derivative), and Robin, which is a linear combination of the Dirichlet and Neumann conditions.\cite{Mazumder_2016} The two currently relevant to our study are Dirichlet and Neumann.

In the Dirichlet boundary condition, the ends of the system's domain are set to a fixed value.  Because the governing equation (Eq.\ \ref{eq:heat-flow-pde-2}) describes temperature differences, zero is the most common fixed value at the boundaries, meaning the ends of the rod terminate in infinite thermal sinks which allow for no temperature change:
\begin{equation} \label{eq:BC_dir} 
    \Theta_0(t) = \Theta_N(t) =0.
\end{equation}

To experimentally achieve the boundary conditions listed in Eq.\ \ref{eq:BC_dir}, we attached two large identical aluminum cylinders to the ends of the copper rod.  If the heat capacities of the aluminum cylinders are much larger than than of the copper rod, then the temperature of the aluminum sinks should be relatively constant and equal to the ambient room temperature.  We refer to this setup as ``sunk" ends of the rod.

In the Neumann boundary condition the ends are defined by a constant flux or change across the boundary.  The most common constant flux is zero, meaning the rod terminates with nothing connected to the ends of the rods, such that no energy is allowed to flow out of the rod:
\begin{equation} \label{eq:BC_neu} 
    \frac{\partial \Theta_0(t)}{\partial z} = \frac{\partial \Theta_N(t)}{\partial z} =0.
\end{equation} 
Implementing this boundary condition requires us to change the second order derivative in space.  The finite centered difference method (Eq.\ \ref{eq:FCD}) has an associated error of order $(\Delta z)^2$.  If the boundary condition is to also have an error that is second order in $\Delta z$, then the Neumann boundary conditions become:\cite{Mazumder_2016}
\begin{align}
   \Theta_0(t+\Delta t) &= \Theta_0(t) + \left(\frac{k(8 \Theta _{1}(t) - \Theta _{2}(t) - 7\Theta _{0}(t) )}{2(\Delta z)^2} + v_0 - w\Theta _0(t)\right)\cdot\frac{\Delta t}{s}, \\  
   \Theta_N(t+\Delta t) &= \Theta_N(t) + \left(\frac{k(8 \Theta _{N-1}(t) - \Theta _{N-2}(t) - 7\Theta _{N}(t) )}{2(\Delta z)^2} + v_N - w\Theta _N(t)\right)\cdot\frac{\Delta t}{s}.  
\end{align}

To achieve these boundary conditions experimentally, we simply attach nothing to the ends of the rod.  We refer to this setup as ``floating" ends of the rods.

We programmed our numerical simulation in Python in two different ways.  Our first method uses the forward Euler method with a fixed time step.  The smaller the time step, the more accurate the simulation.  Our second method uses the built-in numerical integration routine \verb"solve_ivp" standard in the \verb"scipy.integrate" library. \verb"solve_ivp" uses a Runge-Kutta integration method (default is the RK45 method) and varies the time step.   We typically divide the rod into $\approx500$ segments, and we used a time step of $\Delta t = 0.001$ s when using Euler method code.  These parameters were chosen by increasing the number of segments and decreasing the time step until the numerical solution no longer changed.  Our code using \verb"solve_ivp" typically takes less than 60 s per simulation to run (the fixed time step code typically takes two to five times longer).  All results presented in this manuscript used \verb"solve_ivp".\cite{EPAPS}

\section{Analysis}

We present experiments on short (22 cm long) rods where we varied the duration of the heat pulse.  We chose to vary the duration of the heat pulse because it was easy to test, had minimal material cost, and was easily repeatable.

\subsection{Evaluating the Numerical Solution}
\begin{figure}
    \includegraphics[width=0.9\linewidth]{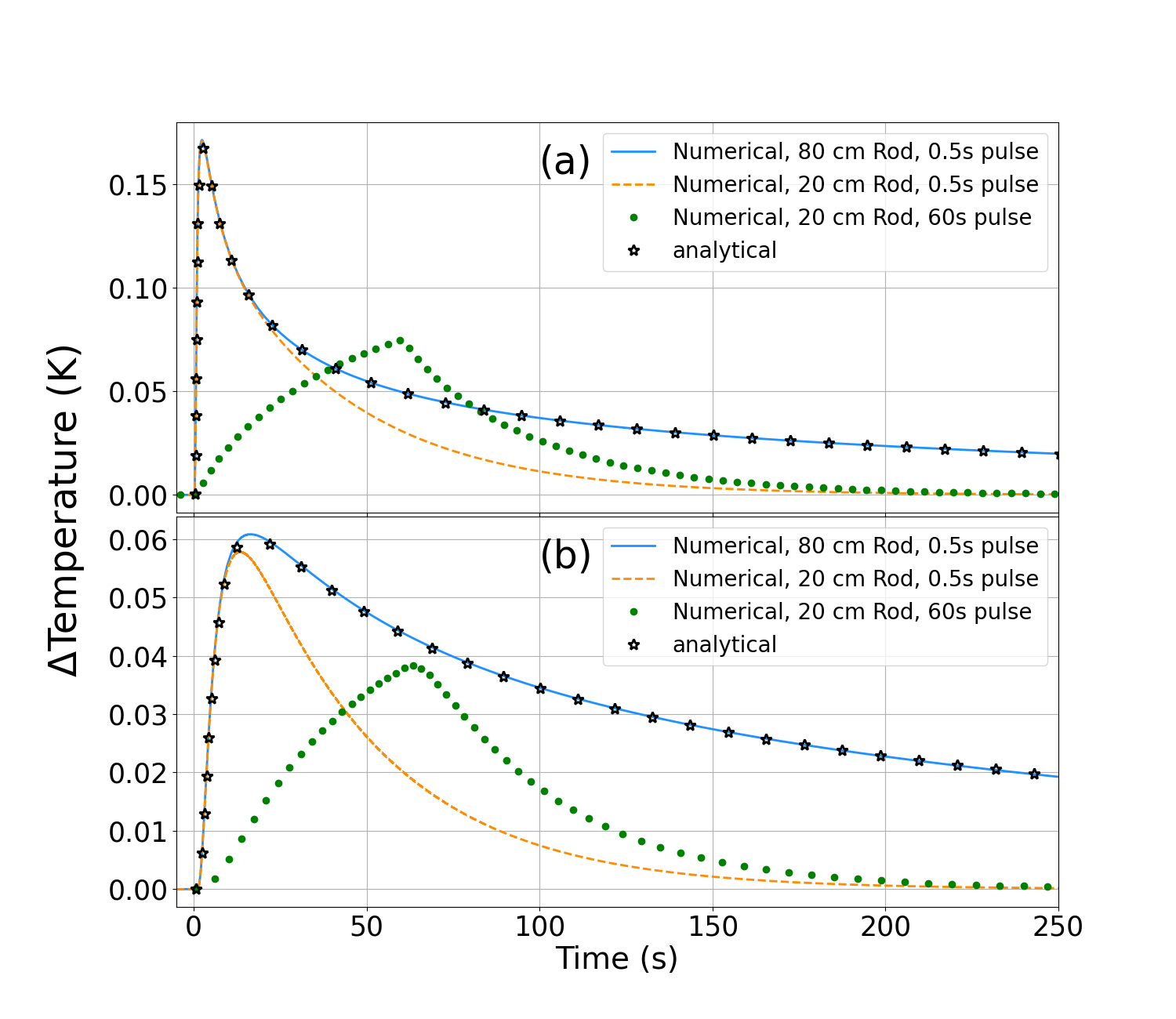}
    \caption{ Numerical and analytical solutions of a $Q= 0.45$ J heat pulse on a copper rod with a 1 cm long heater at (a) $z = 2$ cm and (b) $z=6$ cm from the center of the rod.  The length of the rod and the duration of the heat pulse is varied.  For long (80 cm) rods and short (0.5 s) heat pulses, the numerical solution matches the analytical solution (stars).}
    \label{fig:analyticalvsnumerical}
\end{figure}

We can use the the analytical solution on an infinite rod (Eq.\ \ref{eq:T-soln2}) to test our numerical solution before deploying the numerical model in novel circumstances.  Figure \ref{fig:analyticalvsnumerical} shows the temperature difference calculated using the numerical model with Dirichlet boundary conditions (Eq.\ \ref{eq:BC_dir})  for three different possible experiments at two different positions away from the heater.  The blue line shows the simulated results for a thermally sunk, 80 cm long rod with a 1 cm long heater.  The $Q= 0.45$ J heat pulse occurs over only 0.5 s, much smaller than the characteristic time $\tau = 3$ s calculated at $z = 2$ cm.  These conditions are sufficiently close to the assumptions of the infinite rod model that the analytical solution (stars) matches the numerical solution, as expected. 

If we break one of the analytical solution's assumptions, for example by reducing the rod length to 20 cm, we can see how the numerical solution (orange line) at first matches the analytical solution, then deviates from it at longer times.  As shown in Fig.\ \ref{fig:analyticalvsnumerical}, the orange line falls back to $\Theta = 0$ much faster than the infinite rod for both the near and far thermometers.  This matches our expectations: it is easier for the energy to ``escape" out of the shorter 20 cm rod and into the aluminum sinks, and thus happens in a shorter time span.

The green dotted line is the numerical solution when we break a second analytical assumption and increase the length of the heat pulse to 60 s (but keep the total energy input $Q = 0.45$ J constant).  In this case, the temperature change peaks at a lower temperature and at a later time.  A longer heat pulse with the same amount of heat allows heat diffusion to begin while heat is still being added. This also matches our expectations: adding the same amount of heat slowly over time will reduce the temperature change in the rod compared to a more rapid addition of heat. Because the heat is added over such a long time, the temperature difference now peaks at nearly the same time for both thermometers.  These results now differ significantly from the results of the analytical solution for an infinite rod.  But do these numerical solutions accurately represent experimental results?

\subsection{Verifying Boundary Conditions} \label{Al_sinks}

\begin{figure}
    \includegraphics[width=0.8\linewidth]{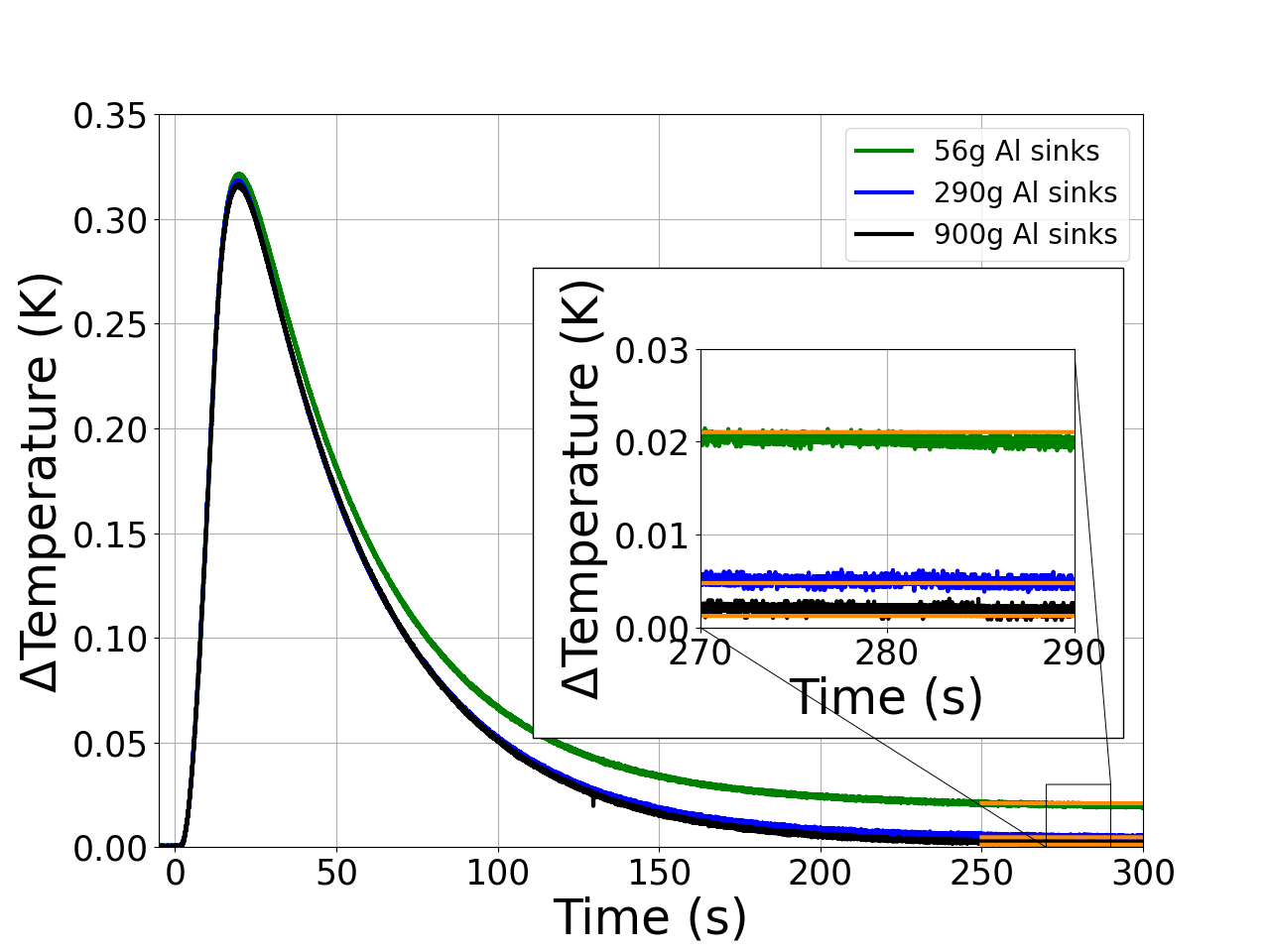}
    \caption{(Color online) A $Q= 2.45$ J, 10 s heat pulse on a 22 cm rod with a 1 cm long heater with three different sets of aluminum sinks at the ends of the rods.  These data were collected in an evacuated bell jar with a thermometer at $z=6$ cm.  For all measurements, the temperature change reaches the value predicted from the thermodynamic limit $Q = m c\Delta T$, given by the orange line, which serves as a check on the experimental setup.  Only for the largest heat sinks (900 g) does the temperature reach a reasonably small value ($\Delta T \lesssim 1$ mK). }
    \label{fig:3}
\end{figure}

Cylindrical aluminum blocks were attached to the copper rod to approximate the infinite thermal sinks used in Eq.\ \ref{eq:BC_dir}. Realistically, the whole system is still floating, since the apparatus is inside an evacuated bell jar and is supported by nested glass beakers with low thermal conductivity.  We started with smaller 56 g aluminum sinks and quickly discovered that 56 g is not large enough to absorb all the energy from the heat pulse without changing temperature significantly.  As shown in Fig.\ \ref{fig:3}, a $Q=2.45$ J heat pulse leads to an equilibrium temperature difference of 0.02 K.  Increasing the mass of the sinks to 290 g each still leads to a temperature difference of $\Delta T \approx 5$ mK at long times.

We can predict this temperature difference using introductory thermodynamics.  From Sec.\ \ref{sec:DiffusionModel} we know the heat loss for measurements taken inside an evacuated bell jar should be very small ($h \approx 1$ Wm$^{-2}$ K$^{-1}$).   In this case, we can use the formula for heat capacity in thermal equilibrium, $Q = mc \Delta T$, to predict the temperature change.  Including the copper rod and the two aluminum sinks, we know: 
\begin{equation}
      \Delta T =  \frac{Q}{(2m_{Al} \cdot c_{Al}) + (m_{Cu} \cdot c_{Cu})} 
      \label{eq:qemcdt}
\end{equation}

The temperature differences at long times for different size thermal sinks are presented in  Fig.\ \ref{fig:3}.  The inset shows the behavior near thermodynamic equilibrium.  The orange lines are the temperature changes predicted using Eq.\ \ref{eq:qemcdt}, which show good fit compared to the experimental data.  These results act as an excellent check on the experimental setup and require no numerical simulation to verify.

In the end, we settled on 900 g aluminum sinks.  These sinks gave a reasonably small temperature difference, $\Delta T \lesssim 1$ mK, and increasing the size of the sinks further would have been difficult to fit within our bell jar apparatus, as the 900 g sinks were already 7.6 cm in diameter and 7.6 cm long.

\subsection{Experiments on short rods and long heat pulses}
We compare our numerical simulations to experimental data collected on a 22 cm long rod, with a heater length of 1 cm and 900 g aluminum heat sinks.  We vary the duration of the heat pulse and consider thermometers at 4 different locations.

\begin{figure}
    \includegraphics[width=0.9\linewidth]{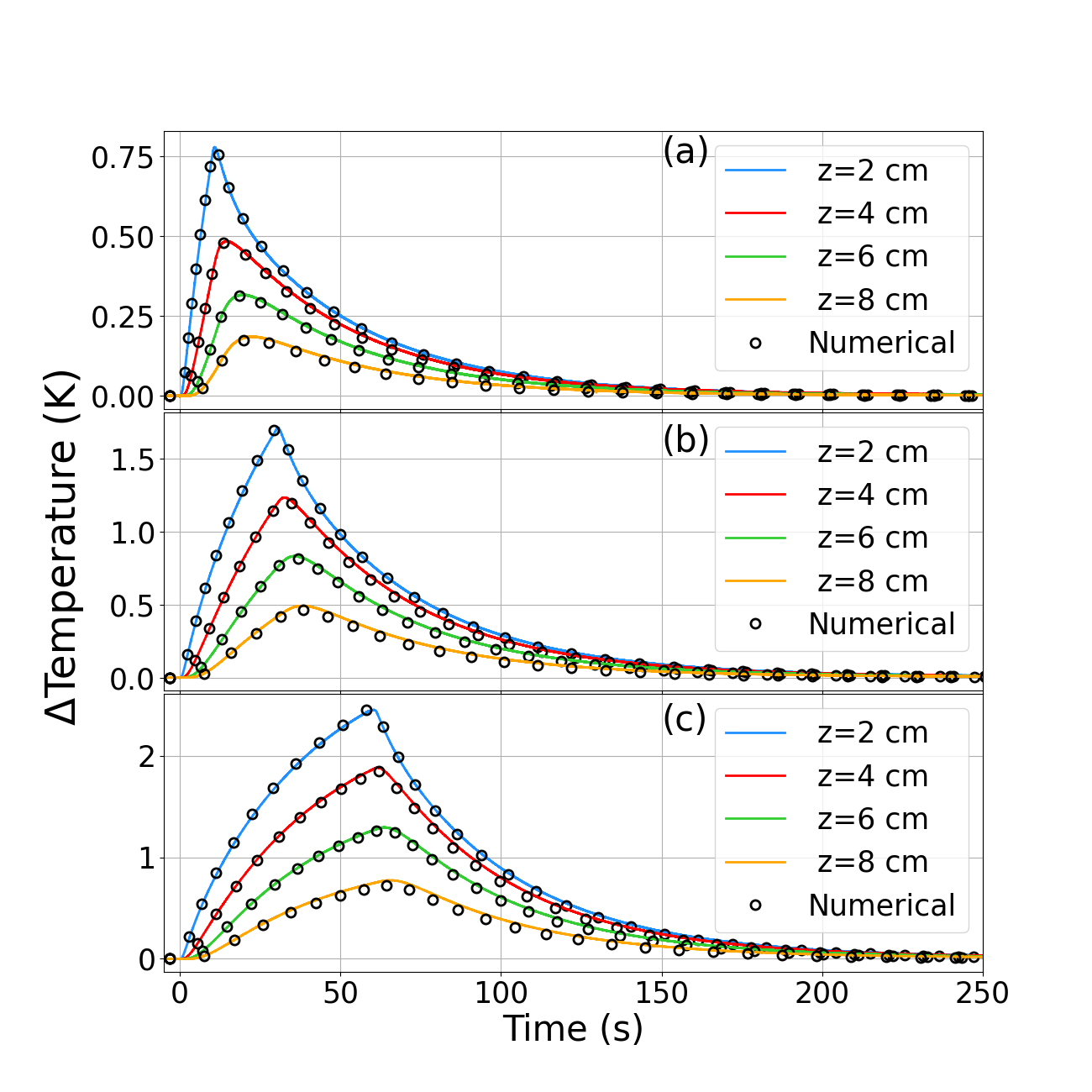}
    \caption{(Color online) Comparison of experimental data and numerical simulation from on a 22 cm long rod with a 1 cm long heater for heat pulses of three different durations: (a) 10 s ($Q= 2.4$ J), (b) 30 s ($Q= 7.2$ J), and (c) 80 s ($Q= 19.3$ J).  These data were collected in an evacuated bell jar.   Each panel shows data from four different thermometers, at $z=2$, 4, 6, and 8 cm.  In all cases, the numerical model very accurately predicts the experimental outcome.}
    \label{fig:experimentalvsnumerical}
\end{figure}

In Fig.\ \ref{fig:experimentalvsnumerical}, we can see the effects of increasing the duration of the heat pulse from 10 s (panel a), to 30 s (panel b), and finally 80 s (panel c), holding the power constant so that the heat added increases from $Q= 2.4$ J to $Q= 7.2$ J and finally $Q= 19.3$ J.  The experimental data for all distances along the rod ($z=2$, 4, 6, and 8 cm) are shown as solid lines.  The numerical simulation matching each heat input and distance is plotted as open circles.  We observe good qualitative agreement between the experimental data and the numerical simulation over a wide range of heat pulse durations and distances along the rod.

For the numerical solutions, we used the model in Eq.\ \ref{eq:FCD} with the Dirichlet boundary conditions in Eq.\ \ref{eq:BC_dir}.  A small effective distance was added to each thermometer distance, $z_{\mathrm{eff}} = 3$ mm.\cite{Sullivan_AJP_2008}   We qualitatively justify this effective distance as the extra distance the heat must travel radially from the surface where the heat is applied to the center of the rod,\cite{Sullivan_AJP_2008} the thermal response time of the thermistors (0.5 s in air), imperfect thermal contact between the rod and the thermistor, imperfect thermal contact between and between the rod and the heater, or  combination of all of these effects.  Furthermore, numerical solutions do not match the analytical solutions if we set the heat loss parameter $h=0$.  For the best fits, we use $h = 3.0$ Wm$^{-2}$ K$^{-1}$.  This heat loss is of the same order of magnitude as the expected heat loss due to radiation, though larger, and probably includes some heat loss due to conduction from the wires and from the connection between the rod and the plastic holder.  Both the effective distance $z_{\mathrm{eff}}$ and the heat loss $h$ were found by minimizing the sum of the square of the residuals between the numerical model and the experimental data simultaneously for all thermometers.  The same $z_{\mathrm{eff}}$ and $h$ parameters were used for all fits presented in this work.

Using $h = 3.0$ Wm$^{-2}$ K$^{-1}$ and adding 3 mm to each thermometer distance for all numerical simulations yield very good agreement between the experimental results and the numerical solution, as shown in Fig.\ref{fig:experimentalvsnumerical}, with only the 8 cm thermometer showing noticeable deviation (less than 30 mK) from the experimental data.  These results work well in an advanced lab environment, when agreements between prediction and experiment are very satisfying to students.  The inclusion of computational physics is essential, as the analytical solution for the infinite rod deviates significantly from the experimental results (Fig.\ \ref{fig:analyticalvsnumerical}).  The numerical solution, on the other hand, provides students with a qualitatively pleasingly good fit in Fig.\ \ref{fig:experimentalvsnumerical}, and requires only two fit parameters for all four thermometers and all times ($z_{\mathrm{eff}}$ and $h$).

\subsection{Failure of the Numerical Model}
Physicists create models of physical systems and use those models to predict experimental outcomes.  As discussed above, agreement between the model and the experiment is usually the preferred outcome, especially in an advanced lab setting.  However, more often than not, the failure of the model to correctly predict the outcome is more interesting and more exciting.  In the advanced lab, this failure is often due to an experimental error, and perhaps just as often blamed on the equipment, but the failure of a model to correctly predict an outcome can lead to a discussion of the limitations of the model and hopefully a new model that better predicts experimental outcomes.\cite{Sullivan_AJP_2022}

\begin{figure}
    \includegraphics[width=0.8\linewidth]{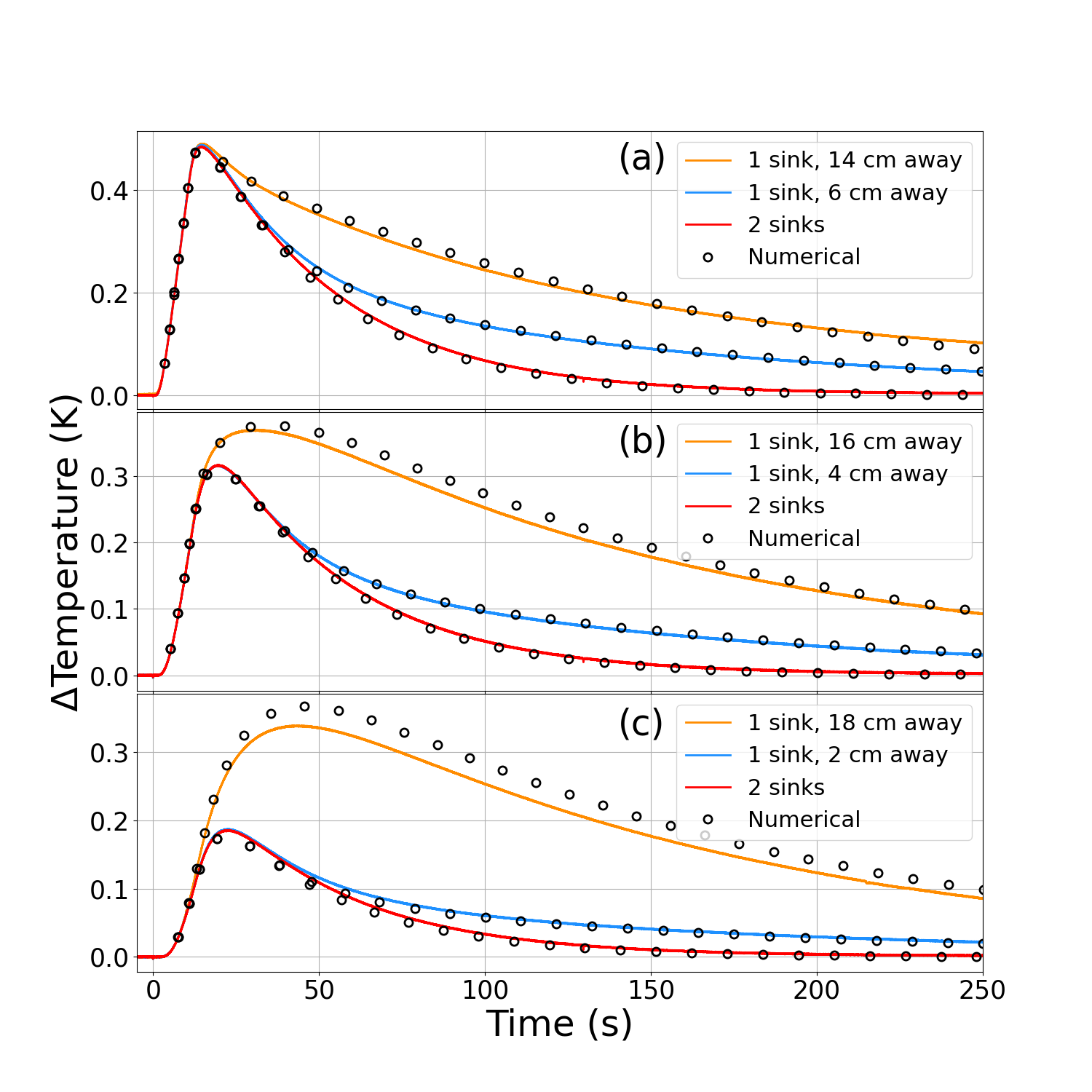}
    \caption{(Color online) The temperature difference due to a $Q= 2.4$ J, 10 s heat pulse on a 22 cm rod with a 1 cm long heater measured at a distance of (a) $z = 4$ cm, (b) $z = 6$ cm, and (c) $z = 8$ cm from the center of the rod. These data were taken in an evacuated bell jar.  Experimental results for aluminum sinks at both ends (red), and for measurements with only one end sunk, with the sink either near the thermometer (blue) or far from the thermometer (orange).  Numerical simulations are open circles.  As before, the simulation matches the experimental data well for sinks at both ends and fits reasonably well when the sink is only 6, 4, or 2 cm away.  When the sink is 14, 16, or 18 cm away, the fit degrades considerably, indicating a failure in the model to capture the dynamics.  }
    \label{fig:FS}
\end{figure}

In this experiment, we can change the boundary conditions in our numerical model and run the simulation with the ends of the rods sunk (Eq.\ \ref{eq:BC_dir}), floating (Eq.\ \ref{eq:BC_neu}), or one end sunk and one end floating.  We can also test this experimentally by simply removing an aluminum sink from one end or both ends of the rod.  In Fig.\ \ref{fig:FS} we present data for a 10 s, $Q=2.4$ J heat pulse in a 22 cm long rod with temperature change measured at a distance of $z = 4$, 6, and 8 cm from the center of the rod for three different configurations: Two heat sinks (the same conditions as in Fig.\ \ref{fig:experimentalvsnumerical}) and one end sunk and one end floating, with the sink switching from the near side to the far side.

For sinks at both ends of the rods, the agreement between the experimental data and numerical simulation remains good (these data are the same as in Fig.\ \ref{fig:experimentalvsnumerical}).  When one sink is removed but the remaining sink is on the near side (6, 4, or 2 cm away), the agreement is not as good, but still convincing.  When we move the sink to the opposite end of the rod (14, 16, or 18 cm away from the thermometer), the agreement between the experiment and the simulation degrades noticeably.  The model developed to describe thermal diffusion now fails to accurately describe the temperature change on the rod.

This experiment in thermal diffusion was motivated by the failure of the analytical model to predict experimental outcomes if the assumptions that led to the analytical solution in Eq.\ \ref{eq:T-soln2} were violated.  That failure led us to look for a new model and a numerical solution to the temperature change on the rod.  In Fig. \ref{fig:FS}, we have reached the limit of the ability of our new model to accurately describe the observed experimental results. Rather than be frustrated, we see this as an opportunity to explore new models or new experiments to explain the observed results.

\section{Conclusion and Challenges}
Computational physics has grown in importance over the last thirty years, and with that growth has become embedded into the undergraduate physics curriculum.  However, there are surprisingly few undergraduate advanced labs that integrate both experiment and numerical simulations.  

This article serves to fill in that gap with an experiment in thermal diffusion whose results can be well modeled numerically but not analytically.  With a total cost well below \$200, our experiment consists of a single 0.32 cm diameter copper rod and a 3D printed plastic support with large aluminum sinks at the end.  This system can be modeled using a simple one-dimensional diffusion equation and the finite centered difference method.  This numerical simulation accurately predicts the temperature change in the rod for heat pulses that vary in magnitude and duration.  This experiment is straightforward without any ``tricky" physics, but provides an excellent experience in machining, CAD drawings, 3D printing, data collection, and numerical modeling.

There are at least two expansions to the experiment that could be used as further challenges.  Students could examine the agreement between the model and the experiment in shorter (and longer) rods and for heaters of different lengths.  Beyond that, we challenge readers to explore the disagreement between the numerical model and the experimental data when one side of the rod is floating.  This is an opportunity to explore revisions to our simple one-dimensional model or modifications to the experimental setup.  Possible routes for further study include: Modeling the thermal contact between the aluminum sink and the copper rod, modeling the conductive heat loss through the electrical wires and the plastic supports, polishing the copper rod to further reduce radiative heat loss, and changing to a three-dimensional thermal diffusion model.


\begin{acknowledgments}
The authors would like to thank Jamie Woodworth for starting the work on this topic and for the first efforts to compare the numerical simulation and experimental data of thermal diffusion in copper rods.  We also thank our colleagues Jerome Fung and Kelley D.\ Sullivan for their careful reading and editing of this manuscript.  The authors would also like to thank the editors and the anonymous reviewers of the manuscript for their careful edits and corrections, which have greatly improved this paper.

 The authors have no conflicts of interest to disclose.  Y.M.\ designed the frame and enclosure and 3D printed it, wrote the numerical simulation, collected and analyzed all the data, wrote the first draft of this manuscript, and rendered all the figures.  M.C.S.\ built the experimental apparatus, assisted in understanding the data, and edited the manuscript.

 \end{acknowledgments}


\bibliography{thermo}

\end{document}